\documentclass{pasj01}
\onecolumn
\draft
\Received{$\langle$reception date$\rangle$}
\Accepted{$\langle$acception date$\rangle$}
\Published{$\langle$publication date$\rangle$}

\newcommand{\fermi}{\textit{Fermi}}
\newcommand{\gr}{$\gamma$-ray}
\newcommand{\snr}{SN~1006}
\newcommand{\uG}{\, \mu{\rm G}} 

\begin{document}

\title{\textit{Fermi} LAT detection of the supernova remnant SN 1006 
revisited: the south-west limb}
\author{Yi Xing\altaffilmark{1,*}, Zhongxiang Wang\altaffilmark{1},
Xiao Zhang\altaffilmark{2}, \& Yang Chen\altaffilmark{2,3}}

\altaffiltext{1}{\footnotesize 
Key Laboratory for Research in Galaxies and Cosmology,
Shanghai Astronomical Observatory, Chinese Academy of Sciences,
80 Nandan Road, Shanghai 200030, China}

\altaffiltext{2}{\footnotesize Department Astronomy, Nanjing University,
163 Xianlin Avenue, Nanjing 210023, China}

\altaffiltext{3}{\footnotesize 
Key Laboratory of Modern Astronomy and Astrophysics,
Nanjing University, Ministry of Education, Nanjing 210023, China}
\email{yixing@shao.ac.cn}


\maketitle

\begin{abstract}
The data from the Large Area Telescope (LAT) onboard the {\it Fermi Gamma-ray
Space Telescope} have recently been updated. We thus re-analyze the LAT data
for the supernova remnant (SNR) SN 1006. Two parts of $\gamma$-ray emission
from the region is clearly resolved, which correspond to the north-east (NE)
and south-west (SW) limbs of the SNR. The former was detected in the previous 
LAT data \citep{xing+16}, but the latter is newly detected in this work. 
The detection of the two limbs are at a $\sim 4\sigma$ significance level, and 
the spectral results for the NE limb is consistent with those obtained
in previous detection analyses. We construct the broadband spectral energy
distribution (SED) for the SW limb. Different scenarios are considered
for the SED in $\gamma$-ray energies. We conclude very similar to
that of the NE limb, the high-energy
and very high-energy emission from the SW limb is likely dominated by
the leptonic process, in which high-energy electrons accelerated from the
shell region of the SNR inverse-Compton scatter background photons to
$\gamma$-rays.
\end{abstract}

\section{Introduction}

As the remnant of the supernova AD 1006, one of 
a few supernovae historically recorded \citep{sg02}, \snr\ 
appears like a disk with a 
diameter of 30\arcmin.
The Galactic latitude of the source is $\sim$14\fdg5, far away from 
the Galactic plane.  
A source distance of 2.2 kpc was derived for \snr\ based on the measurements 
of the proper motion of the shock front and the expanding velocity obtained 
with optical observations \citep{gha+02,wgl03}. 
It was the first supernova remnant (SNR) with a 
non-thermal X-ray emission component detected \citep{koy+95}, 
as the radio and hard X-ray emission from the front shell regions is 
dominated by synchrotron emission (e.g., \cite{rg86,rot+04,win+14}), while
the interior region was found to have a thermal spectrum with line 
features (e.g., \cite{uyk13}).
The X-rays emitted by \snr\ indicate that the electrons can be accelerated 
to 100 TeV energies in the shock front \citep{koy+95}. 
Therefore it is considered as an efficient site of cosmic rays 
acceleration in 
the Milky Way, although in a relatively low ambient-density environment.

\begin{figure}
\begin{center}
\includegraphics[scale=0.25]{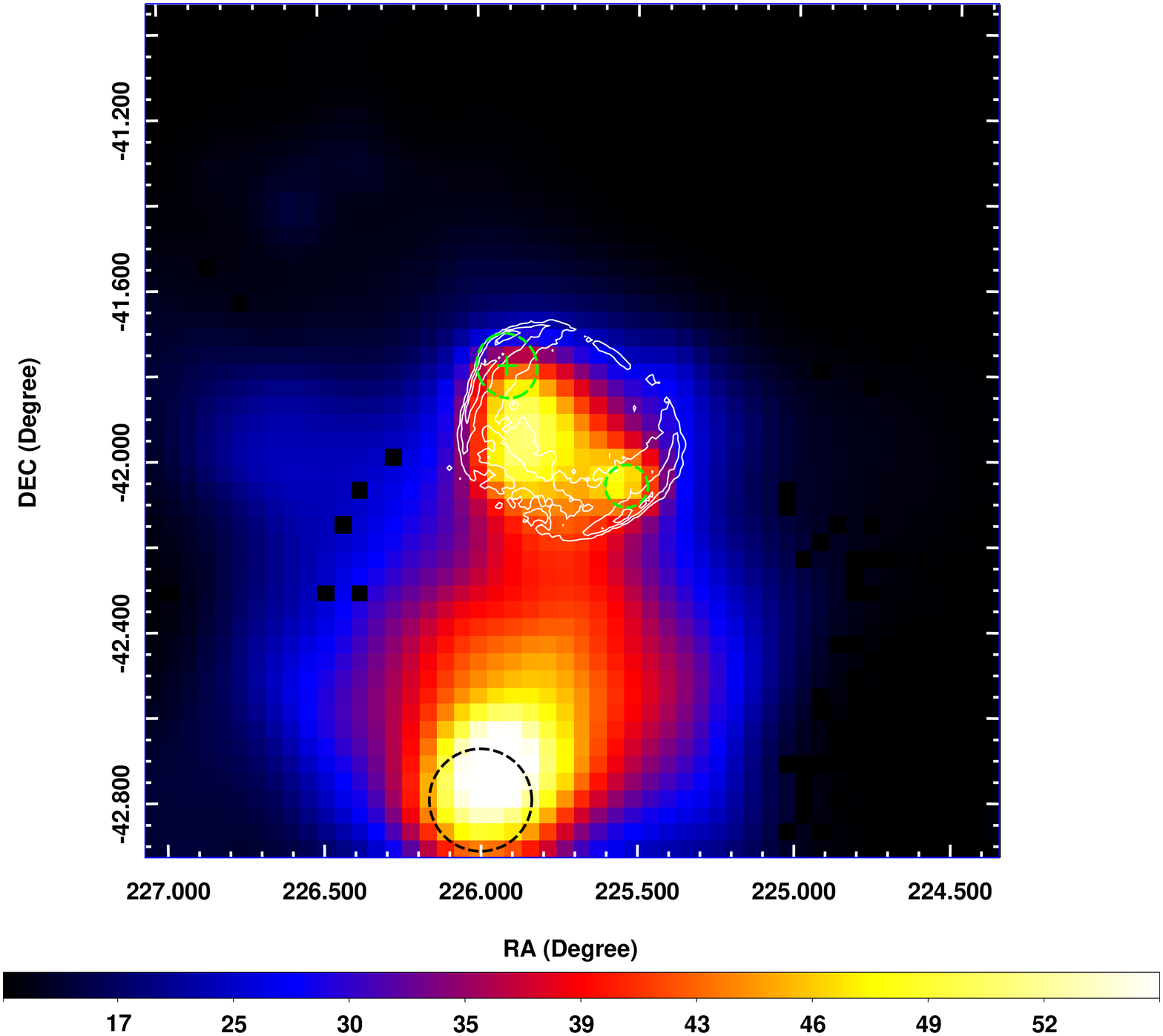}
\includegraphics[scale=0.25]{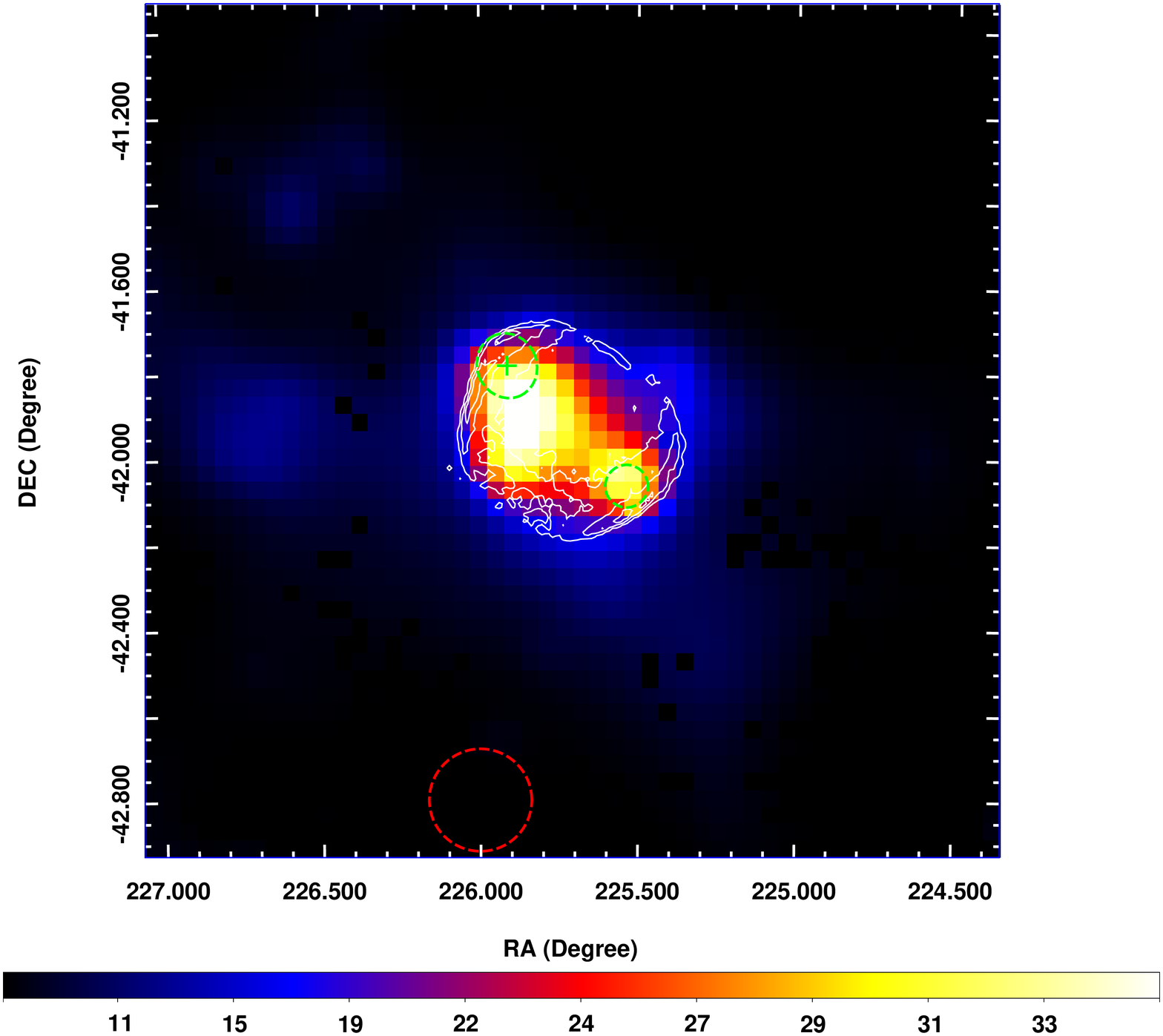}
\end{center}
\caption{TS maps of the $\mathrm{2^{o}\times2^{o}}$ region centered 
at \snr\ in the 0.5--500 GeV band.  The image scale of the maps is 
0\fdg04 pixel$^{-1}$. All catalog sources except the counterpart to the NE 
limb were considered and removed.  
The green plus and ellipse mark the position and the 2$\sigma$ error ellipse of the NE limb given in the 4FGL catalog, and
the green circle is the 2$\sigma$ error circle of the best-fit position 
obtained for the SW limb. The white contours indicate the 0.75--1.3 keV X-ray 
intensity measurements of \snr.
The dark or red circle marks the 2$\sigma$ error circle of 
the best-fit position obtained for 
a nearby source, which is revealed and removed in 
the {\it left} and {\it right} panel, respectively.}\label{fig:tsmap}
\end{figure}

As part of the shell of the SNR, two identifiable limbs are at 
northeast (NE) and southwest (SW) regions (cf., Figure\ref{fig:tsmap}). 
In the very high-energy (VHE; $>$100 GeV) ranges, two sources
were detected with
the High Energy Stereoscopic System (HESS) as HESS J1504$-$418 
and HESS J1502$-$421, which correspond to the NE and SW limbs,
respectively \citep{ace+10}. In the high-energy range of 0.1--300 GeV, 
the detection of the NE limb was reported 
by \citet{xing+16} at a $\sim$4$\sigma$ significance level from
analysis of the data obtained with the Large Area Telescope (LAT) onboard the 
{\it Fermi Gamma-ray Space Telescope (Fermi)}. 
Although the GeV counterpart was not well resolved due to the large 
point-spread function of the LAT and the low detection significance, 
the detection was confirmed by \citet{con+17} at a $\sim$6$\sigma$ 
significance level.
In any case, the broadband spectral energy distribution (SED) of the NE limb 
was found to be well described with a leptonic scenario, in which 
the high-energy and VHE photons are produced
from the inverse Compton (IC) process of high-energy electrons \citep{xing+16}.
The synchrotron emission of the limb is from the same population of the 
electrons.

While the GeV \gr\ emission from the SW limb has not been detected in 
the previous studies, \citet{mic+14} reported evidence for the interaction 
with a HI cloud in the region, making it to a promising region 
for \gr\ hadronic emission in \snr. 
In the HESS VHE observation, the brightness of the NE limb
is $\sim$50\% higher than that of the SW limb.
Given these, it was predicted by \citet{xing+16} that the SW limb
would likely be detected with the accumulation of 
the \fermi\ LAT data. 
With the release of the updated \fermi\ LAT Pass 8 data (P8R3) and 
the accumulation of $>$10 years of data (comparing to 7 years and
8 years in \cite{xing+16} and \cite{con+17}, respectively),
re-analysis of the LAT data for \snr\ is warranted.
In this paper, we report the results from our analysis of the data.
Now GeV \gr\ emission from both the NE and SW limbs are detected.
We describe the \fermi\ LAT data and source model for analysis in Section~2,
and present the data analysis and results in Section~3.
Different models are discussed in Section~4 to explain the SED of the SW limb.

\section{\textit{Fermi} LAT Data and Source Model}

LAT is one of the two main instruments onboard \fermi, conducting
all-sky survey in the energy range from below 20 MeV to more than 300 GeV \citep{atw+09}.
For this analysis, we selected the 0.1--500 GeV LAT events 
from the recently updated \textit{Fermi} P8R3
database. The region considered is $\mathrm{20^{o}\times20^{o}}$, with
the central position  
at the radio center of \snr\ (R.A.=15$^{\rm h}$02$^{\rm m}$50, 
Decl.=$-$41$^{\circ}$56$'$00, equinox J2000.0). The time period of the data
selected was from 2008-08-04 15:43:36 (UTC) to 2018-12-02 08:29:55 (UTC), more
than 10 years.
As suggested by the LAT 
team\footnote{\footnotesize http://fermi.gsfc.nasa.gov/ssc/data/analysis/scitools/}, 
events with zenith angles greater than 90 degrees and with quality flags 
of `bad' were excluded, which are for the purpose of preventing
the Earth's limb contamination and the spacecraft events affection, 
respectively.

We included sources within 20 degrees centered at the position of 
\snr\ to make the source model. The LAT 8-year point source catalog\footnote{\footnotesize https://fermi.gsfc.nasa.gov/ssc/data/access/lat/8yr\_catalog/} 
(4FGL) was recently released in early 2019, which was used in the analysis to extract the source model.
The spectral forms of these sources are provided 
in the catalog. 
The background Galactic and extragalactic diffuse 
emission models used were  
gll\_iem\_v07.fits and the file iso\_P8R3\_SOURCE\_V2\_v1.txt, respectively.

\section{Data Analysis and Results} 
\label{sec:dar}

\subsection{Maximum Likelihood Analysis}
\label{subsec:mla}

We performed the standard binned likelihood analysis to the LAT data 
in the $>$0.5 GeV band using {\tt Fermitools 1.0.0}. 
The low-energy data in the $<$0.5 GeV band were not used, which
helps to reduce the 
effects of the Galactic background and to
avoid the relatively large uncertainties 
of the instrument response function of the LAT in the low energy range. 
The spectral parameters of the sources 
within 5 degrees from \snr\ and the normalizations of the diffuse components were set as free parameters, and 
the other parameters of the sources were fixed at their catalog values.
The NE limb has already been included in the source model as a point source 
4FGL J1503.6$-$4146, with power-law emission. 
From the analysis, we found photon index $\Gamma= 2.3\pm$0.2 
and 0.5--500 GeV flux $F_{0.5-500}= 5.9\pm1.7\times 10^{-10}$ photons~s$^{-1}$\,cm$^{-2}$ for 
the NE part. The photon index value is consistent with those previously
reported within uncertainties \citep{xing+16,con+17}.

Using the fitted source model, we constructed a 0.5--500 GeV Test Statistic 
(TS) map of a $\mathrm{2^{o}\times2^{o}}$ region centered at the radio position
of \snr. All of the known sources given
in the LAT catalogs were removed from the TS map, except the \gr\ counterpart 
of the NE limb.
The obtained TS map is shown in the left panel
of Figure~\ref{fig:tsmap}. The \snr\ region is
resolved as two parts in the TS map. Besides the NE limb (TS$\sim 50$), 
which previously appeared slightly extended (but could not be 
determined from likelihood analysis;
\cite{xing+16,con+17}) covering most of the SNR's disk, now there is isolated
excess emission at the SW limb with TS$\sim 45$.

However in the bottom of Figure~\ref{fig:tsmap}, there is an extra source 
with TS$\sim 60$. In order to check whether it might affect our detection,
we ran {\tt gtfindsrc} in {\tt Fermitools} to determine its position:
R.A.=226\fdg00, Decl.=$-$42\fdg80, 
(equinox J2000.0), with 1$\sigma$ nominal uncertainty of 0\fdg08.
Considering it as a point source with power-law emission in the source model,
we re-performed the likelihood analysis. The resulting TS map with this source
removed is shown in the right panel of Figure~\ref{fig:tsmap}. 
The two parts of excess emission at the NE and SW limbs can still be clearly
revealed, with slightly lower TS values of $\sim$35 and $\sim$30, respectively.
We ran {\tt gtfindsrc} to determine the position of the SW part and obtained 
R.A.=225\fdg54, Decl.=$-$42\fdg06, 
(equinox J2000.0), with 1$\sigma$ nominal uncertainty of 0\fdg03.

Finally we included both of the NE and SW parts in the source model as point 
sources at the catalog position and the obtained best-fit position, 
respectively, 
and re-performed the likelihood analysis. The source, $\sim0\fdg8$ away from
\snr\ in the south (the bottom of Figure~\ref{fig:tsmap}), was also 
included in the source model. 
We obtained $\Gamma= 1.7\pm$0.2 
and $F_{0.5-500}= 1.6\pm0.7\times 10^{-10}$ photons~s$^{-1}$\,cm$^{-2}$,
for the NE part, and $\Gamma= 2.0\pm$0.2 
and $F_{0.5-500}= 2.0\pm1.0\times 10^{-10}$ photons~s$^{-1}$\,cm$^{-2}$ 
for the SW part. 
Based on the face values, the \gr\ emission from the NE limb seems to be 
harder than that from the SW limb, but the uncertainties are large and
no conclusion can be made.
The TS values for the NE and SW parts are 20 and 23, respectively, 
both corresponding to $>$4$\sigma$ detection significance. 

\subsection{Spatial Distribution Analysis}
\label{subsec:sda}

Since the \gr\ emission from both of the NE and SW limbs of \snr\ are 
detected by \fermi, we performed the likelihood analysis with the total 
excess emission in the SNR region considered to be one extended source.
From this analysis, we examined whether the excess emission would be better 
described as one extended 
source or two separate point sources. A template was created with 
the {\it XMM-Newton} image (0.75--1.3 keV)\footnote{\footnotesize https://heasarc.gsfc.nasa.gov/docs/xmm/gallery/esas-gallery/xmm\_gal\_science\_snr1006.html}, shown
as the white contours in Figure~\ref{fig:tsmap}. A power-law emission model 
was assumed for the extended source. We obtained $\Gamma= 2.0\pm$0.2 
and $F_{0.5-500}= 6\pm2\times 10^{-10}$ photons~s$^{-1}$\,cm$^{-2}$,
with a TS value of 46. The TS value indicates a $>$6$\sigma$ detection 
significance. 
However, the likelihood value for the extended source model ($L_{e}$) is 
approximately equal to that for the two point sources model ($L_{2ps}$) in
our analysis, indicating that the former is not preferred based on the current
data. Considering that the two limbs may have different emission origins, 
in the following sections we treated the excess \gr\ emission as two separate 
point sources. 
\begin{figure}
\begin{center}
\includegraphics[scale=0.25]{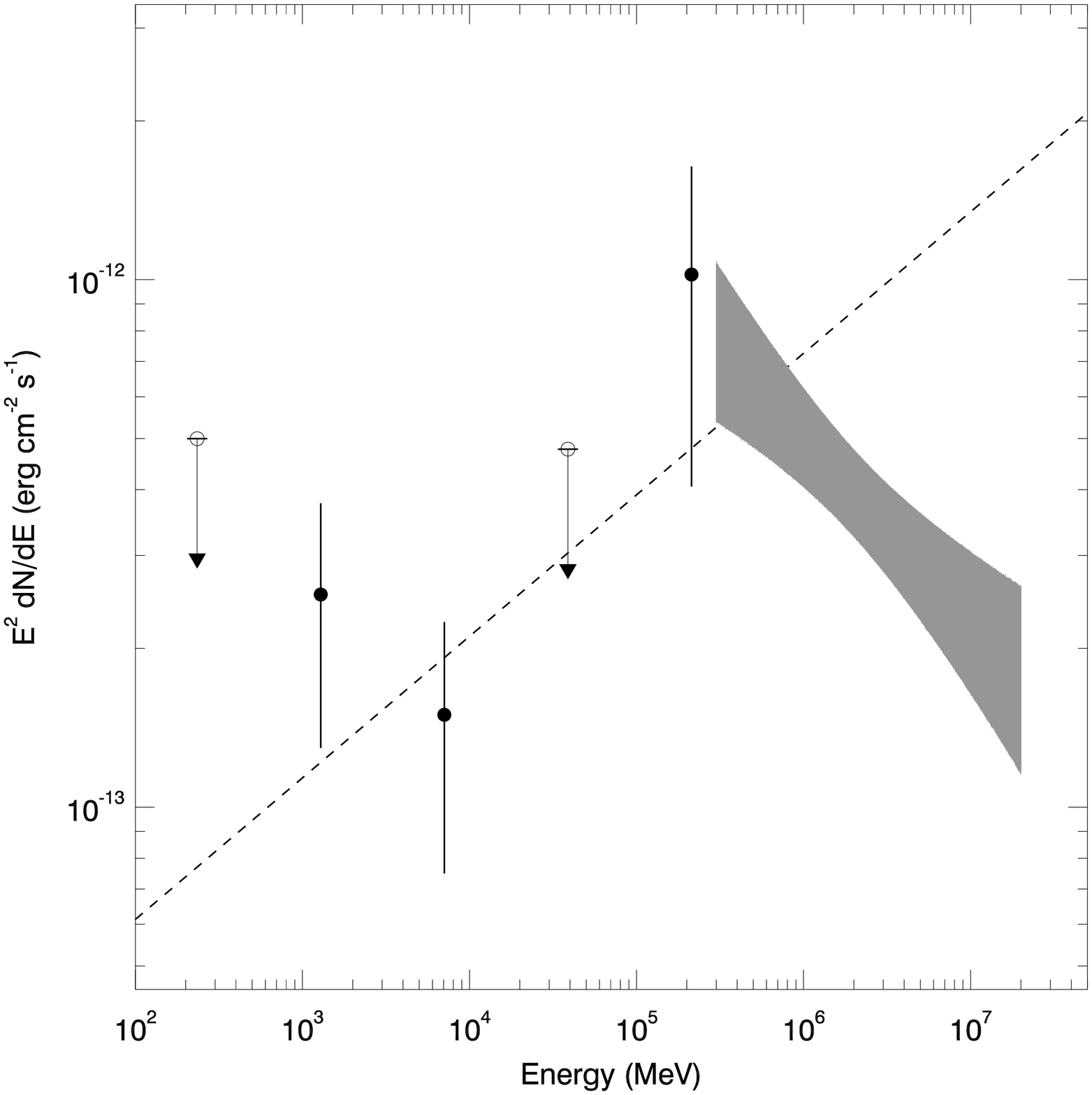}
\includegraphics[scale=0.25]{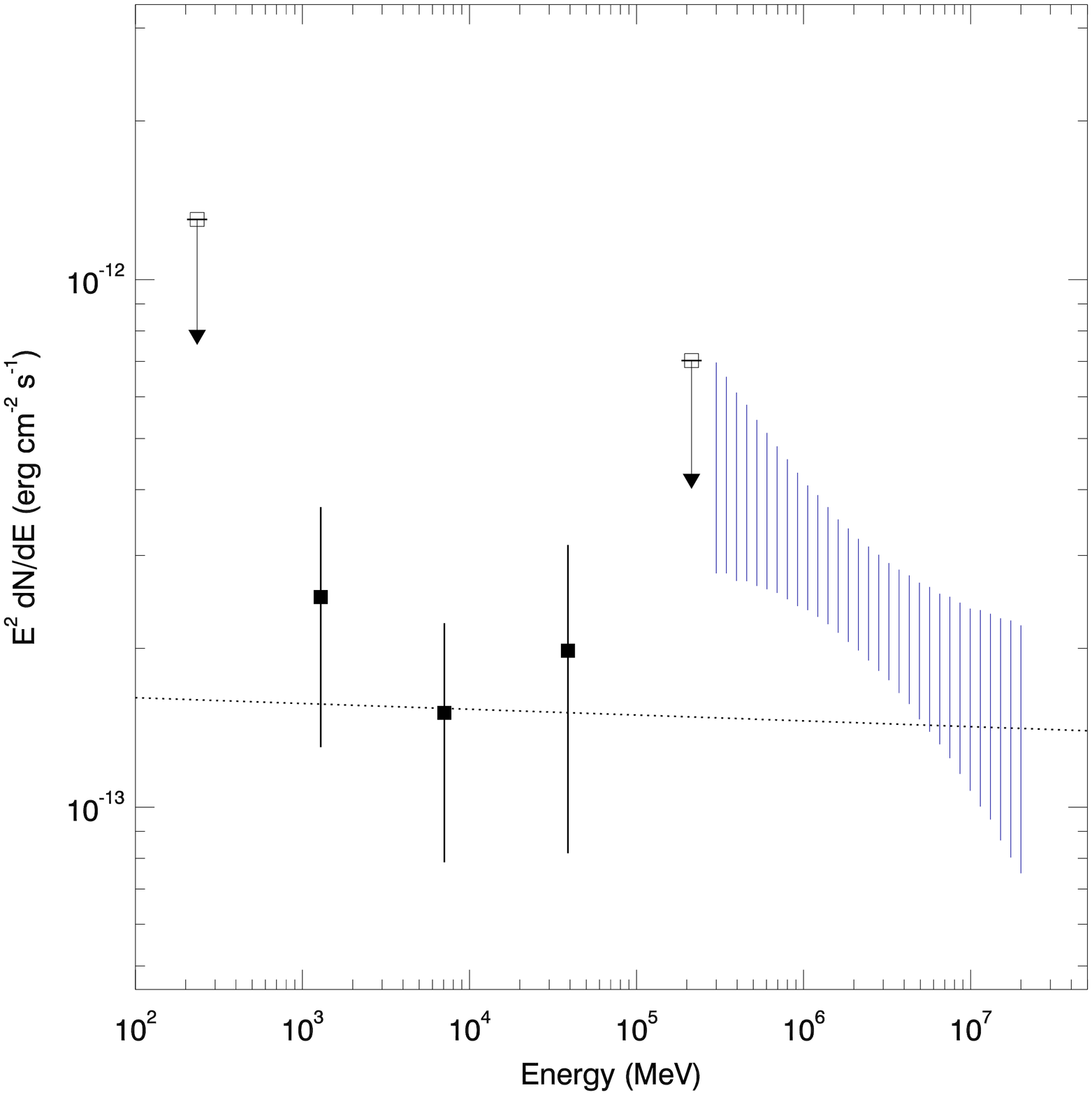}
\end{center}
\caption{\fermi-LAT \gr\ spectra and the 0.5--500 GeV power-law fits of 
the NE ({\it left}) and SW ({\it right}) limbs
of \snr. The gray and blue areas mark the HESS power-law spectra of 
the NE and SW limbs of SN 1006, respectively.}\label{fig:spectra}
\end{figure}

\subsection{Spectral Analysis}
\label{subsec:sa}

We extracted the $\gamma$-ray spectra of the NE and SW parts by including 
them as point sources with power-law emission and performing maximum 
likelihood analysis to the LAT data 
in 5 evenly divided energy bands in logarithm from 0.1--500 GeV.
In the extraction, the spectral normalizations of the 
sources within 5 degrees from the central position of \snr\ were set as 
free parameters, 
while all the other parameters of the sources were fixed at the values 
obtained from the above maximum likelihood analysis.
We kept only spectral flux points with the TS values greater than 5 ($>$2$\sigma$), 
or derived 95\% flux upper limits otherwise. 
The obtained spectra are shown in 
Figure~\ref{fig:spectra}, with
the fluxes and uncertainties provided in Table~\ref{tab:spectra}.
  
\subsection{Variability Analysis}
\label{subsec:lv}

In order to fully study the \gr\ emission properties of \snr, we also 
searched for any long-term variability of it. We calculated the variability 
index TS$_{var}$ for the NE and SW parts with 126 time bins 
(each bin was constructed from 30-day data) in the energy ranges 
of 0.5--500 GeV, following the procedure 
introduced in \citet{nol+12}. If the flux is constant, 
TS$_{var}$ would be distributed as $\chi^{2}$ with 125 degrees of freedom. 
Variable sources would be identified with TS$_{var}$ larger than 164.7 
(at a 99\% confidence level).
The computed TS$_{var}$ for the NE and SW parts are 64.3 and 65.3, 
respectively, indicating that there were no significant 
long-term variability in them. 
\begin{figure}
\begin{center}
\includegraphics[scale=0.65]{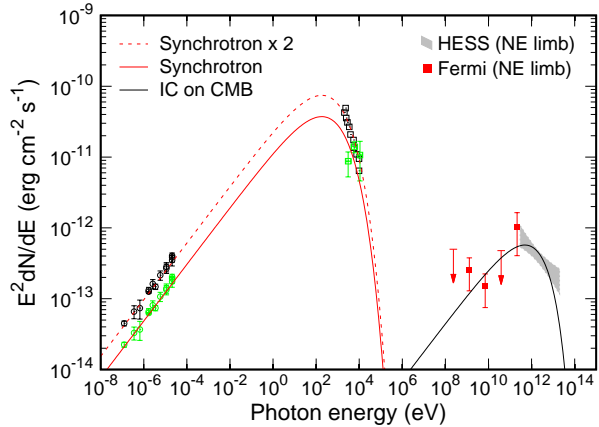}
\end{center}
\caption{Leptonic model fit (red and black curves) to the broadband SED 
of the NE limb of SN 1006. 
Radio \citep{apg01}, X-ray \citep{bam+08}, and 
HESS \citep{ace+10} data for SN~1006 are plotted as dark open circles, 
dark squares, and gray region, respectively. The green squares are X-ray 
measurements for the NE part given in \citet{kal+06}, and green open circles 
are half of the radio fluxes obtained for the whole remnant.}
\label{fig:sed_ne}
\end{figure}

\section{Discussion}
\label{sec:disc}

Having analyzed $>$10 years of \fermi\ LAT P8R3 data, the excess \gr\ emission 
at the \snr\ region is clearly resolved, with two parts located at the
NE and SW limbs. The obtained spectral parameters for
the previously detected NE limb are consistent with those reported
in \citet{xing+16} and \citet{con+17}. However the detection significance is 
$\sim$4$\sigma$, lower than that obtained with 8 years of data 
(6$\sigma$) in \citet{con+17}. This issue is likely due to the use of 
4FGL and the updated Pass 8 data. Within 5 degrees from \snr,
there are 12 sources in 4FGL but only 5 sources in 3FGL, and 
the updated Pass 8 data has a residual background significantly lower than 
that of the previous data. Both these results in a cleaner background 
in the analysis. Likely due to the same reasons, we were able to detect
the SW limb with a significance of $\sim$4$\sigma$, in addition to
the longer time period of the data used.
We also showed that if we considered the total excess emission as 
one extended source with the spatial distribution the same as that 
in the X-ray image, the \gr\ emission, following a $\Gamma= 2.0$ power law, 
could be detected with a $>$6$\sigma$ significance.

For the NE limb, we update its broadband SED in Figure~\ref{fig:sed_ne},
by including the spectral data points obtained in Section~\ref{subsec:sa}.
Comparing to that in \citet{xing+16}, the radio \citep{apg01},
X-ray \citep{bam+08}, and HESS TeV \citep{ace+10} measurements remain 
unchanged. In addition, the X-ray flux measurements for the NE limb given
in \citet{kal+06} are added. Since there are no significant changes in
the data points, the leptonic model previously used in \citet{xing+16} 
can provide a fit to the updated SED (the electron spectral index $\alpha_{e}$ $=$ 2.2, electron cutoff energy $E_{\rm cut,e}\approx17$ TeV, and magnetic field strength of the SNR $B_{\rm SNR}\approx24\uG$; refer to \cite{xing+16} for the detailed model and model parameters). In the model calculation, the synchrotron flux 
was multiplied by a factor of 2, as the radio and X-ray data were 
from the whole SNR and the two limbs were assumed to be symmetric 
for simplicity. The model fit is shown in Figure~\ref{fig:sed_ne} 
(red and black curves), and the model parameters given in \citet{xing+16} were adopted. We note that recently by fitting the radio and {\it XMM-Newton} and {\it NuSTAR} X-ray data for the NE and SW limbs, \citet{li+18} obtained electron spectral indices of $\sim$1.9 and electron cutoff energies of $\sim$7 TeV. These values are different from what we used here. In their work, however, the magnetic field strength was fixed at 100$\uG$ and a single-frequency radio flux measurement was used.

\begin{figure}
\begin{center}
\includegraphics[scale=0.65]{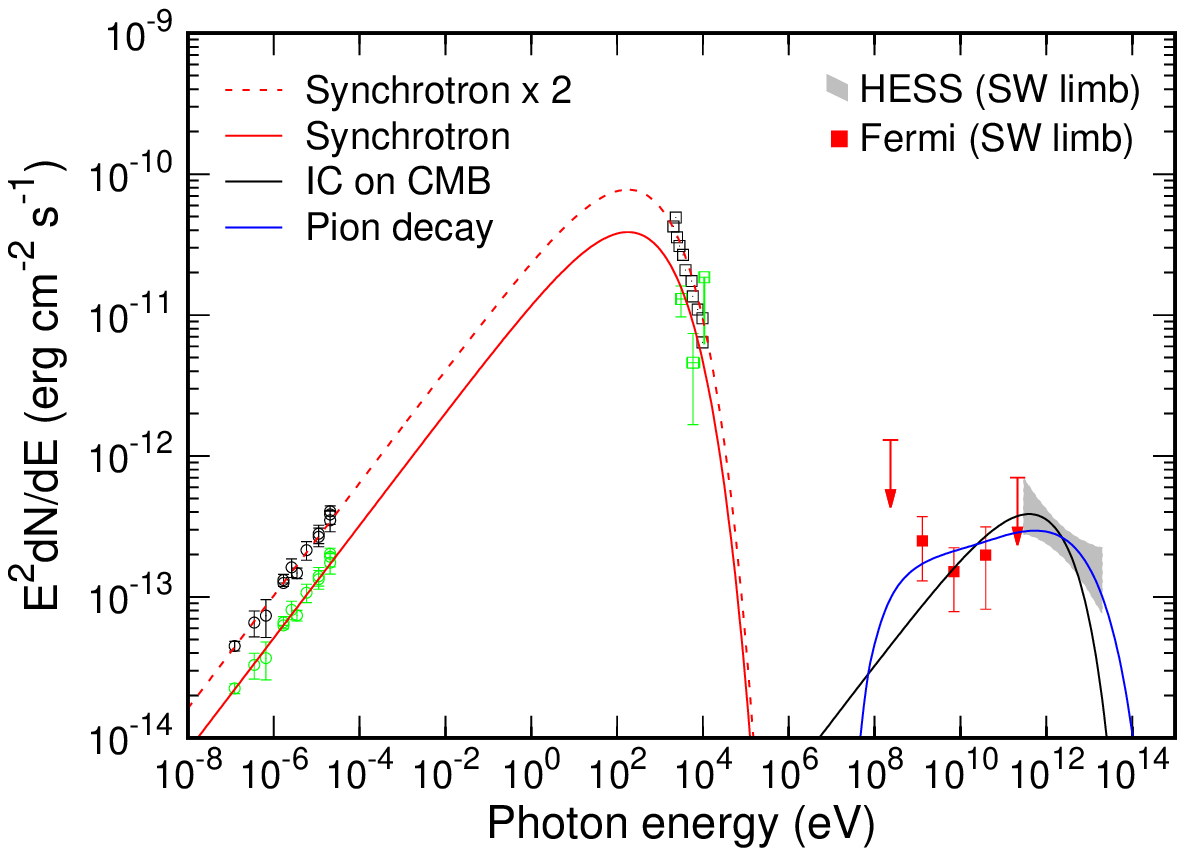}
\includegraphics[scale=0.65]{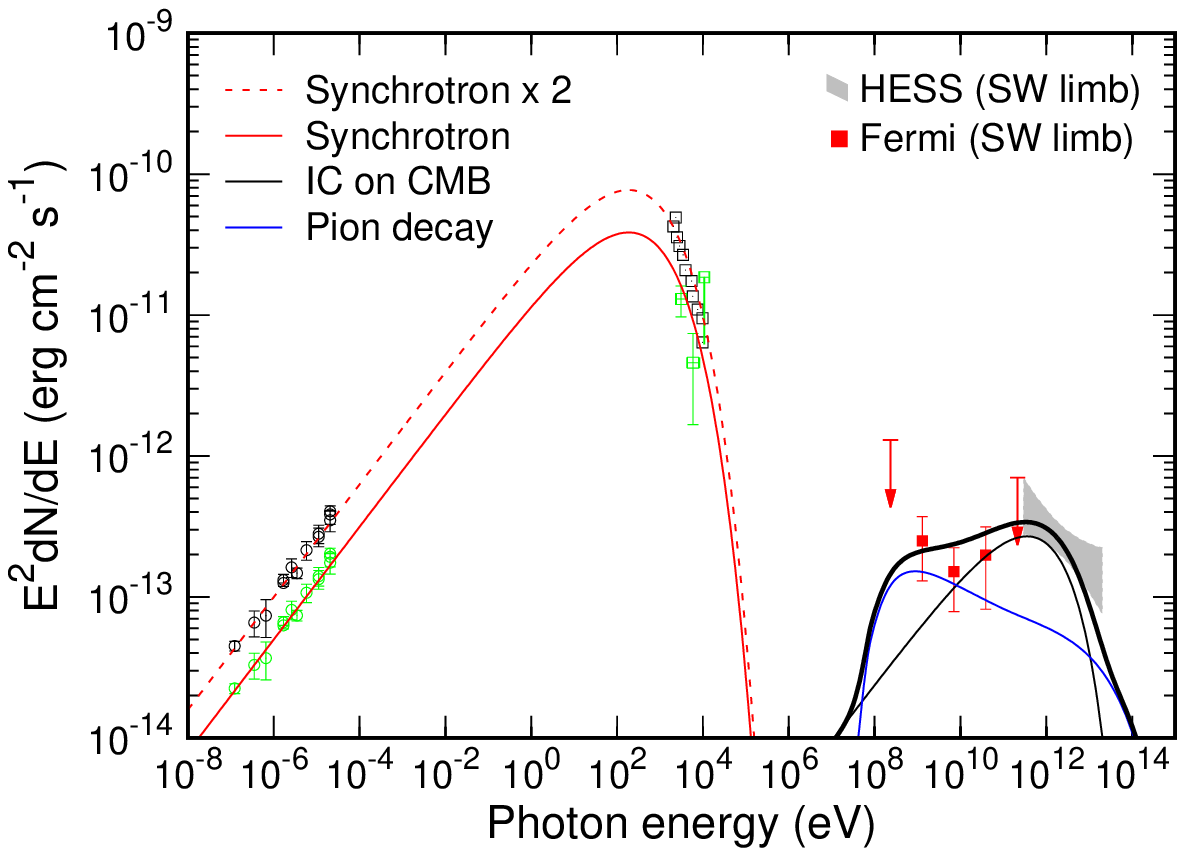}
\end{center}
\caption{Leptonic and hadronic model fits to the broadband SED of 
the SW limb of SN 1006. Radio \citep{apg01}, X-ray \citep{bam+08},
and HESS \citep{ace+10} data are shown as dark open circles, dark squares,
and gray region, respectively. The green squares are the X-ray measurements for 
the SW part given in \citet{kal+06}, and the green circles are half of 
the radio fluxes obtained for the whole remnant. A purely leptonic
(black curve) or hadronic (blue curve) model fit for the \fermi\ and 
HESS \gr\ data is shown in 
the {\it left} panel, and a combined contribution 
(thick black curve) from
the mixed leptonic (black curve) and hadronic (blue curve) processes 
is shown in the {\it right} panel. 
}
\label{fig:sed_sw}
\end{figure}

For the SW limb, we constructed its broadband SED and show it in 
Figure~\ref{fig:sed_sw}. The radio and X-ray data are the same as those 
used for the NE limb, and the HESS TeV measurement is from \citet{ace+10}. 
In addition, the X-ray flux measurements for the SW limb from
\citet{kal+06} are also included.
We first considered the purely leptonic model used for the NE limb and 
found that it can well explain the SED of the SW limb. In deriving
the model fit shown in the left panel of Figure~\ref{fig:sed_sw} 
(red and black curves), the electron cutoff energy 
$E_{\rm cut,e}\approx15$ TeV, the total electron energy of the SW limb
$W_e(>1{\rm GeV})\approx1.1\times10^{47}$ erg, and magnetic field strength
$B_{\rm SNR}\approx30\uG$.
The values are very similar to those used for the NE limb 
(see \cite{xing+16}).

It has been shown in \citet{ace+10} that the HESS TeV emission 
cannot be attributed to a purely hadronic model, and when taking into account
the \fermi\ LAT measurements at the time, the hadronic origin was ruled out 
at a $>$5$\sigma$ confidence level in \citet{ace+15_tevsnr} assuming an extended source hypothesis and a \gr\ flux fixed to the hadronic hypothesis.
It can be noted that the tenuous environment around \snr, with
the density of the interstellar medium (ISM) estimated to be
$n_{\rm ism}\sim$0.035 cm$^{-3}$ \citep{mic+16},
does not favor the proton-proton interactions.
However, \citet{mic+14} reported the detection of a dense HI cloud interacting 
with the SW limb of \snr, which suggests a high ambient density for 
the hadronic process in this region. 
We thus also considered the hadronic origin for \gr\ emission of the SW limb. 

There are two hadronic components, one is from the shocked cloud, and the other is from the shocked ISM.
For simplicity, we assumed that their parent hadrons have the same energy 
distribution function of 
$dN_{p}/dE_{p}\propto E_{p}^{-\alpha_{p}} {\rm exp}(-E_{p}/E_{cut,p})$.
In a purely hadronic scenario, 
we obtained the proton spectral index $\alpha_p\approx1.95$, the
cutoff energy $E_{cut,p}\approx150$ TeV, and the
total proton energy 
$W_p(>1{\rm GeV})\approx2.5\times10^{49}(n/0.2\ {\rm cm^{-3}})^{-1}$ erg, 
where $n$ is the average density of the gas in downstream. 
Considering $n_{\rm ism}$ $=$ 0.035 cm$^{-3}$, the average cloud density
$n_{\rm cloud}$ $=$ 0.5 cm$^{-3}$ \citep{mic+16}, and a proper geometric 
factor of 1/64 (here we considered half of the remnanet; \cite{mic+14}),
$n$ is approximately 0.2\,cm$^{-3}$. 
Although the model can fit the \gr\ SED (the blue curve in the left 
panel of Figure~\ref{fig:sed_sw}), $\alpha_{p}\approx1.95$ is less than 
2.2 derived from the radio data and the canonical value of 2.0 (predicted 
by the standard diffusive shock acceleration theory).
Therefore \gr\ emission from the SW limb is not likely hadronic-dominated, 
which is consistent with that suggested in \citet{ace+10} 
and \citet{ace+15_tevsnr}. 

A mixed scenario that includes 
leptonic and hadronic components may be considered. We fixed 
$\alpha_{p}$ $=$ 2.2, the same as that of the electrons. A model fit
shown in the right panel of Figure~\ref{fig:sed_sw} (the thick black 
curve) can be obtained. The leptonic component (the black curve) 
has $E_{\rm cut,e}\approx14$ TeV, $W_e(>1{\rm GeV})\approx0.8\times10^{47}$ erg, 
and $B_{\rm SNR}\approx36\uG$.
The hadronic component 
(the blue curve) has $E_{cut,p}$ $=$ 1 PeV and 
$W_p(>1{\rm GeV})\approx1.5\times10^{49}(n/0.2\ {\rm cm^{-3}})^{-1}$ erg.
Assuming a canonical explosion energy of 10$^{51}$\,erg and a symmetric 
geometry, the energy value implies that the fraction of explosion energy 
converted into hadrons for the full remnant is lower than 
the order of $\sim$2\% for the current stage. 
In this mixed scenario, the contribution of the leptonic component is only 
$<$30\% lower than that of the purely leptonic model. 
The \gr\ emission from the SW limb is still leptonic-dominated.

In summary, our analysis of the recently updated \fermi\ Pass 8 data for
the \snr\ region results in the \gr\ detection of not only the NE part 
reported in \citet{xing+16} and \citet{con+17}, but also the SW part that
has not been detected in previous studies. 
The 0.5--500 GeV \gr\ luminosity of the NE and SW parts are
$1\times 10^{33}$ erg~s$^{-1}$ and $6\times 10^{32}$ erg~s$^{-1}$, respectively.
Similar to the emission process considered for the NE part, we find that
the \gr\ emission of the SW part 
likely arises from the leptonic process with reasonable model parameters.
The luminosity values support our modeling, as they are
two orders of magnitude lower than those of the dynamically evolved SNRs 
\citep[and references therein]{xin+15}. \gr\ emission from the latter group
originates from the hadronic process, due to the interaction between 
the SNRs and nearby molecular clouds.
We suspect that the interacting area between the shock surface of SN 1006 
and the HI cloud is much lower than that estimated in \citet{mic+14} or
that the energy conversion efficiency of the SNR is lower than $\sim$2\% at 
the present age.
Thus the high-energy and VHE $\gamma$-ray emission from the SW limb of 
SN~1006 is dominated by the leptonic process.

\bigskip
This research made use of the High Performance Computing Resource in the Core
Facility for Advanced Research Computing at Shanghai Astronomical Observatory.
This research was supported by the National Program on Key Research 
and Development Project (Grant No. 2016YFA0400804) and
the National Natural Science Foundation
of China (U1738131,11373055). 
X.Z. and Y.C. acknowledge the supports from the National Program on Key Research and Development Projects 2018YFA0404204, 2017YFA0402600, and 2015CB857100, and from the NSFC under grants 11803011, 11773014, and 11851305.

\begin{table}
\tbl{\fermi\ LAT flux measurements of the NE and SW
limbs of \snr.\footnotemark[$*$] }{%
\begin{tabular}{lccccc}
\hline
$E$ & Band & $F/10^{-13}$ (NE) & TS (NE) & $F/10^{-13}$ (SW) & TS (SW) \\
(GeV) & (GeV) & (erg cm$^{-2}$ s$^{-1}$) &  & (erg cm$^{-2}$ s$^{-1}$) & \\
\hline
0.23 & 0.1--0.5 & $<$5.0 & 0 & $<$13.0 & 4 \\
1.29 & 0.5--3.0 & 3$\pm$1 & 6 & 3$\pm$1 & 6 \\
7.07 & 3.0--16.6 & 1.5$\pm$0.7 & 7 & 1.5$\pm$0.7 & 7 \\
38.84 & 16.6--91.0 & $<$4.8 & 3 & 2$\pm$1 & 7 \\
213.34 & 91.0--500.0 & 10$\pm$6 & 7 & $<$7.0 & 0 \\
\hline
\end{tabular}}\label{tab:spectra}
\begin{tabnote}
\footnotemark[$*$]$F$ is the energy flux ($E^{2} dN/dE$).  Fluxes without uncertainties are the 95$\%$ upper limits.
\end{tabnote}
\end{table}

\end{document}